%% file: ConformationMain.tex
\documentclass[10pt,twocolumn, amssymb, nobibnotes, aps, amsmath, amssymb, superscriptaddress,floatfix]{revtex4-2}
\usepackage{graphicx}
\usepackage{dcolumn}
\usepackage{bm}
\usepackage[linktocpage,colorlinks=true,linkcolor=blue,citecolor=blue,breaklinks=true,urlcolor=blue]{hyperref}
\usepackage{float}
\usepackage[utf8]{inputenc}
\usepackage[T1]{fontenc}
\usepackage{newtxtext, newtxmath}  %% these are more "modern" fonts
\DeclareGraphicsExtensions{.pdf,.jpeg,.jpg, .png, .eps, .tiff}
\usepackage{epstopdf}
\usepackage{lipsum}
\usepackage{natbib}
\usepackage[dvipsnames]{xcolor}
\usepackage[normalem]{ulem}
\usepackage{soul}
\usepackage{enumitem}
\usepackage{siunitx}
\usepackage{lineno}
\usepackage{subfigure}
\usepackage{anyfontsize}
\renewcommand{\vec}[1]{\underline{#1}}

\newcommand{\vecP}[2]{{#1}_{#2}}
\makeatletter
\newcommand{\tens}[1]{{\mathpalette\tens@{#1}}}
\newcommand{\tens@}[2]{%
  \begingroup
  \sbox\z@{$\m@th#1\underline{#2}$}%
  \dimen@=\dp\z@ \advance\dimen@ -2\tens@dimen{#1}%
  \dp\z@=\dimen@
  \sbox\z@{$\m@th\underline{\box\z@}$}%
  \box\z@
  \endgroup
}
\newcommand\tens@dimen[1]{%
  \fontdimen8
  \ifx#1\displaystyle\textfont\else
  \ifx#1\textstyle\textfont\else
  \ifx#1\scriptstyle\scriptfont\else
  \scriptscriptfont\fi\fi\fi 3
}
\makeatother
\newcommand{\kbt}{k_\text{B}T}
\newcommand{\VCO}{\Xi}
\newcommand{\OCO}{\Psi_c} 

\newcommand{\av}[1]{\left\langle #1 \right\rangle}
\newcommand{\abs}[1]{\left| #1 \right|}
\newcommand{\Par}[1]{\left( #1 \right)}

\newcommand{\fig}[1]{\textbf{Fig.~\ref{#1}}}
\newcommand{\eq}[1]{\textbf{Eq.~\ref{#1}}}

\newcommand{\sctn}[1]{\textbf{\S~\ref{#1}}}

\newcommand{\EQ}[1]{\textbf{Equation~\ref{#1}}}
\newcommand{\act}{\zeta}

\newcommand{\actlen}{l_\text{a}}

\newcommand{\SystemSize}{l_\text{sys}}

\newcommand{\re}{R_\text{e}}
\newcommand{\reTrans}{R_{\text{e},0}}

\newcommand{\plus}{+1/2}

\newcommand{\ue}{School of Physics and Astronomy, The University of Edinburgh, Peter Guthrie Tait Road, Edinburgh, EH9 3FD, United Kingdom}

\begin{document}

\title{Passive polymers in active turbulence undergo a collapse-stretch transition}

\author{Zahra K. Valei}
\affiliation{\ue}
\author{Davide Marenduzzo}
\affiliation{\ue}
\author{Tyler N. Shendruk}
\email{t.shendruk@ed.ac.uk}
\affiliation{\ue}

\date{\today}

\begin{abstract}

Active processes in living systems generate nonequilibrium forces that deform embedded passive macromolecules.
To understand how such dynamics influence polymer conformation, we study a flexible passive chain in an active nematic fluid. 
Using hybrid simulations, we uncover a length-dependent transition in polymer behavior: long chains align with and stretch along defect-driven flows, while short chains bend and collapse due to localized stresses. These responses are controlled by a competition between the polymer size and the emergent length scale of the active turbulence.
Our results reveal a defect-mediated mechanism for conformational control and point toward general physical principles for designing responsive soft materials that couple passive structure to active dynamics.

\end{abstract}

\maketitle
%%%%%%%%%%%%%%%%%%%%%%%%%%%%%%%%%%%%%%%%%%%%%
%%%%%%%%%%%%%%%%%%%%%%%%%%%%%%%%%%%%%%%%%%%%%
%%%%%%%%%%%%%%%%%%%%%%%%%%%%%%%%%%%%%%%%%%%%%
%% INTRODUCTION %%
%%%%%%%%%%%%%%%%%%%%%%%%%%%%%%%%%%%%%%%%%%%%%
%%%%%%%%%%%%%%%%%%%%%%%%%%%%%%%%%%%%%%%%%%%%%
%%%%%%%%%%%%%%%%%%%%%%%%%%%%%%%%%%%%%%%%%%%%%
\section{Introduction}
\label{sctn:intro}
    
The conformational response of polymers to self-organized, persistent motion in active environments remains a fundamentally open problem, with direct relevance to a variety of biological processes.
% option2: to the directed motion and interactions of surrounding active agents — such as self-propelled particles or molecular motors —
% option3: to nonequilibrium forces exerted by surrounding active elements
For example, structural remodeling of chromatin by force-generating proteins--including RNA polymerase and topoisomerases--plays a crucial role in regulating gene transcription~\cite{eshghi2022symmetry,nozaki2023condensed,goychuk2023polymer}. 
Protein-mediated DNA looping is essential for gene regulation in both prokaryotic and eukaryotic organisms~\cite{cherstvy2011looping}.
% Actin filaments and microtubules, as core components of the cytoskeleton, are actively reorganized by molecular motors such as myosin and kinesin. This dynamic machinery forms the cell’s contractile framework, supporting essential processes from cell migration and division to large-scale deformations like muscle contraction
Networks of actin filaments and microtubules are dynamically reorganized by molecular motors, such as myosin and kinesin, giving rise to structures
that drive processes like cell migration, division, and muscle contraction~\cite{flormann2024structure,banerjee2020actin,lee2021active,shamipour2021cytoplasm}.
% cytokinetic ring assembly, like this phrase!
Developing a unifying physical framework is key to linking nonequilibrium conformational dynamics of biopolymers to the diverse mechanical roles they play in living systems.

Numerous models have been developed to capture the effects of active fluctuations on polymers, accounting for different aspects of active forces.
These studies range from dry systems--in which hydrodynamic interactions are screened--to wet systems--which include solvent-mediated flow fields.
In dry systems, activity is typically introduced via self-propelled particles, often modeled as active Brownian particles (ABP)~\cite{harder2014activity,shin2015facilitation,nikola2016active}.
These particles exert steric forces on polymers, generally leading to chain swelling~\cite{kaiser2014unusual}.
To date, explicit ABP–polymer models that include hydrodynamic interactions have not yet been reported. 

\begin{figure*}
    \centering
    \includegraphics[width=\linewidth]{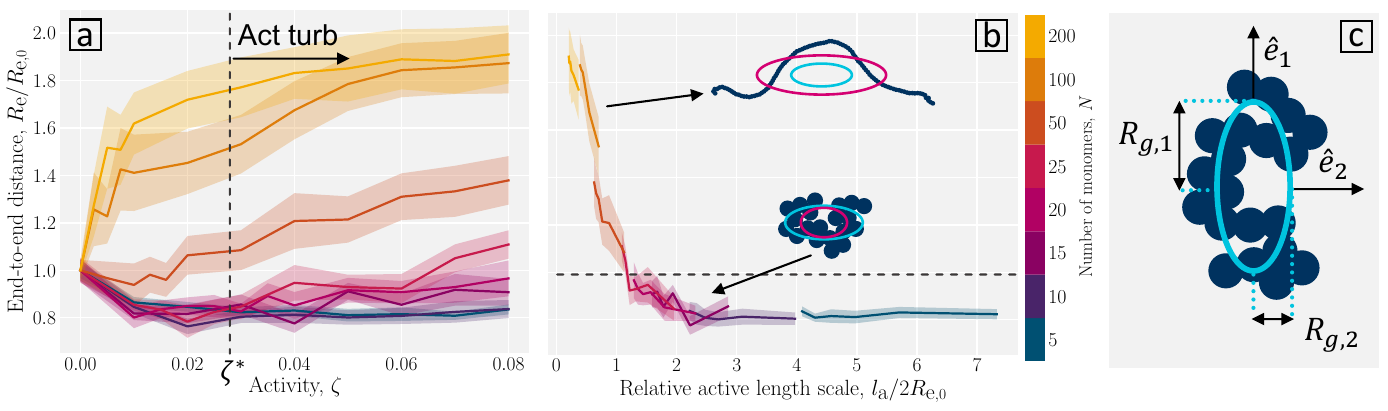}
	\caption{End-to-end distance statistics of polymers in an active nematic background.
    (a)
    End-to-end distance $\re$ as a function of activity $\zeta$,
    normalized by $\reTrans$ is the thermal end-to-end distance measured in a passive medium.
    The vertical dashed line indicates the critical activity $\zeta^* =0.03$, marking the onset of active turbulence, beyond which the flows are characterized by an active length scale $\actlen \sim \zeta^{-1/2}$.
    (b)
    Normalized end-to-end distance $\re/\reTrans$ as a function of the relative active length scale $\actlen/2\reTrans$ for various polymer lengths $N$ in the active turbulence regime ($\zeta \geq 0.03$).
    Insets show representative polymer conformations above and below the coil-stretch transition.
    Ellipses represent the principal moments of the gyration tensor:
    % Ellipses provide a visual summary of the polymer shape and anisotropy:
    blue for polymers in passive fluids and pink for the highest activity $\zeta = 0.08$.
    The smaller polymer has $N=20$ navy-colored monomers of size $\sigma$; the longer has $N=100$ monomers.
    (c)
    Schematic representation of the polymer's gyration tensor.
    The principal axes $\hat{e}_1$ and $\hat{e}_2$ denote the eigenvectors of the gyration tensor, with corresponding principal radii of gyration $R_{\text{g},1}$ and $R_{\text{g},2}$. The light blue ellipse visualizes the anisotropic envelope of the polymer conformation.
    The ellipses in (a) are constructed using the time-averaged values \( \langle R_{\text{g},i} \rangle + \sigma_{R_{\text{g},i}} \), where \( \sigma_{R_{\text{g},i}} \) denotes the standard deviation across time.
    }
	\label{fig:REE}
\end{figure*}

Beyond explicit ABP baths, activity can also be incorporated at a coarse-grained level through effective stochastic forces acting directly on the polymer.
Here, the persistent collisions of active particles are coarse-grained into exponentially correlated active noise, reflecting the finite persistence time of active motion.
This approach has been applied to both partially active~\cite{loi2011non,dutta2024effect,luo2024langevin,shen2025knotting} and fully active chains~\cite{elgeti2015physics,ghosh2014dynamics,samanta2016chain,osmanovic2017dynamics,eisenstecken2017internal}, predicting activity-induced swelling of flexible polymers~\cite{kaiser2015does,vandebroek2015dynamics,martin2018active,chaki2019enhanced,eisenstecken2016conformational}.
In most of these models, the active forces are isotropically oriented, but they can also be aligned with local polarity, acting tangentially along the polymer contour.
Depending on the balance between bending rigidity and propulsion, this can lead to stable spiral conformations~\cite{chelakkot2014flagellar,isele2015self,isele2016dynamics}.
While such coarse-grained descriptions often neglect hydrodynamics, in wet active systems solvent-mediated flows fundamentally alter polymer conformational dynamics.
Activity in such systems may be internally driven--via self-propelled, force- and torque-free monomers--or externally actuated by imposed driving forces.
Externally actuated polymers stretch monotonically with increasing activity~\cite{martin2020hydrodynamics}.
Internally driven polymers, by contrast, exhibit non-monotonic behavior, including shrinkage at intermediate activity and reduced swelling at higher activity~\cite{martin2019active,clopes2022simulating}, highlighting the strong coupling between activity and hydrodynamic flow.
Nevertheless, most existing models neglect the spontaneous, collective flows and mesoscale turbulence that emerge in active fluids, which are key drivers of intracellular mixing and cytoplasmic streaming~\cite{chakrabarti2024cytoplasmic,htet2023cortex,drechsler2020optical,wensink2012meso,htet2025analytical}.
Despite reports of both swelling and shrinkage, a unified understanding that accounts for the polymer’s intrinsic size, the characteristic length scale of activity, and the role of hydrodynamic interactions remains elusive.

We study the conformations of a single passive, freely jointed polymer embedded in active nematics.
Active nematics in bulk exhibit active turbulence, a low-Reynolds number phenomenon with characteristic length and time scales~\cite{shankar2022topological,alert2022active}.
We find that passive polymers embedded in active nematic turbulence undergo activity-induced conformational changes that depend strongly on chain length.
As activity increases, short polymers collapse, long ones stretch, and intermediate lengths show a non-monotonic response.
This composite system exemplifies the competition between entropy-driven relaxation processes within the passive polymer and activity-driven dynamics. 
  
The length-dependent behavior arises from the way in which the polymers interact with nearby $\plus$ defects that structure the active turbulence.
Long polymers align with and stretch due to directional defect-driven flows approaching from the transverse sides and departing along the chain.
In contrast, short polymers exhibit an asymmetric defect distribution, with accumulation near their sides and depletion along their axis, compressing the chain and promoting compaction.

%%%%%%%%%%%%%%%%%%%%%%%%%%%%%%%%%%%%%%%%%%%%%%%%%%%%%%%%%%
%%%%%%%%%%%%%%%%%%%%%%%%%%%%%%%%%%%%%%%%%%%%%%%%%%%%%%%%%%
%% Model and Method
%%%%%%%%%%%%%%%%%%%%%%%%%%%%%%%%%%%%%%%%%%%%%%%%%%%%%%%%%%
%%%%%%%%%%%%%%%%%%%%%%%%%%%%%%%%%%%%%%%%%%%%%%%%%%%%%%%%%%
\section{Numerical model and Method}

We employ a hybrid simulation approach, combining Active Nematic Multi-Particle Collision Dynamics (AN-MPCD) to model the active nematic background and Molecular Dynamics (MD) to simulate the polymer.
The AN-MPCD technique provides a mesoscopic description of active nematic fluids, enabling the simulation of embedded particles within a fluctuating active nematohydrodynamic background~\cite{kozhukhov2022mesoscopic} (Appendix \sctn{sec:AN-MPCD}).
The polymer is represented as a flexible linear chain of $N$ beads, which interact via bonded and excluded-volume potentials and evolve under an MD scheme~\cite{hospital2015molecular} (Appendix \sctn{sec:MD}).
Beads exchange momentum with fluid particles, thereby coupling to the velocity field but not directly to the nematic orientation.
This framework incorporates fluctuations through the thermal energy $\kbt$, hydrodynamic interactions, nematic elasticity via $K$, and active stresses through the activity $\act$.

\begin{figure*}
    \centering
    \includegraphics[width=\textwidth]{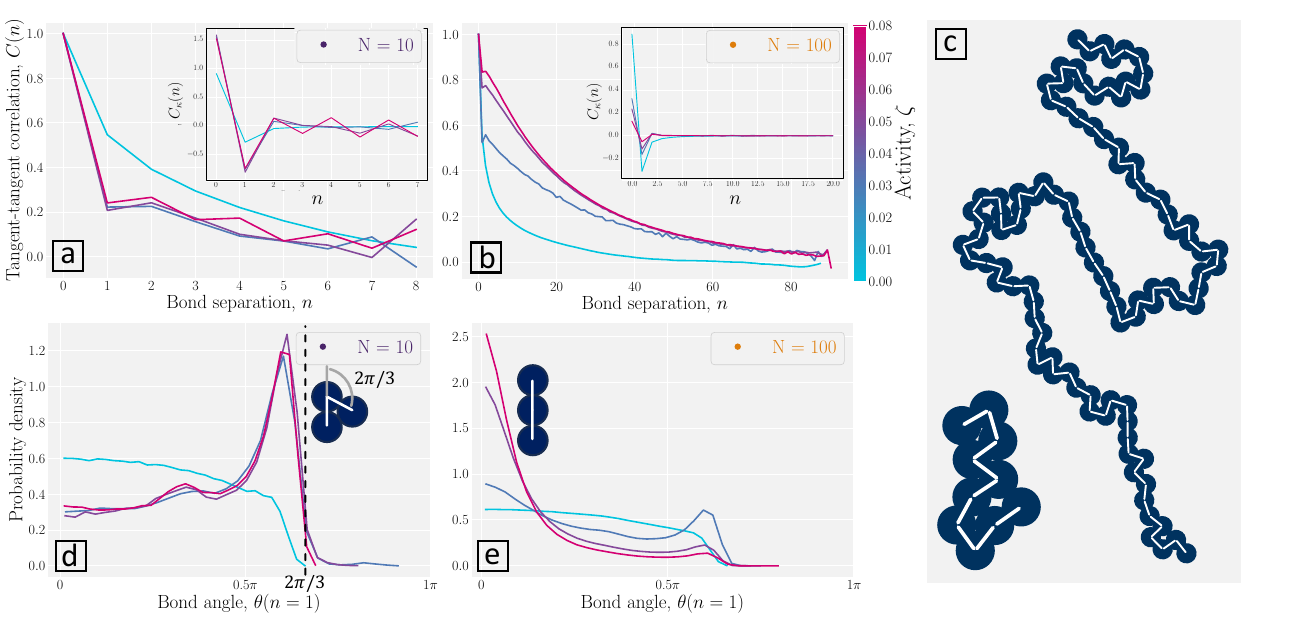}
	\caption{Tangent–tangent correlations, curvature correlations, and bond angle distributions. 
    (a,b) Tangent–tangent correlation functions $C(n)$ for polymers with (a) $N=10$ and (b) $N=100$ monomers at selected activity levels, as in \fig{fig:pdf}.
    Insets show curvature–curvature correlations $C_\kappa(n)$ as a function of bond separation $n$.
    (c) Representative polymer conformations at fixed activity $\zeta = 0.05$, comparing a short chain (bottom left) and a long chain (center).
    (d,e) Probability density functions of the bond angle between consecutive segments $\theta(n=1)$, for polymers with $N=10$ and $N=100$ monomers.
    Insets show the minimum and maximum accessible bond angles imposed by excluded-volume interactions, along with example conformations illustrating compact (d) and extended (e) configurations.
    }
	\label{fig:correlations}
\end{figure*}
%

%%%%%%%%%%%%%%%%%%%%%%%%%%%%%%%%%%%%%%%%%%%%%%%%%%%%%%%%%%%%%
%%%%%%%%%%%%%%%%%%%%%%%%%%%%%%%%%%%%%%%%%%%%%%%%%%%%%%%%%%%%%
%% Results
%%%%%%%%%%%%%%%%%%%%%%%%%%%%%%%%%%%%%%%%%%%%%%%%%%%%%%%%%%%%%
%%%%%%%%%%%%%%%%%%%%%%%%%%%%%%%%%%%%%%%%%%%%%%%%%%%%%%%%%%%%%
\section{Results}
First, we discuss how the conformations of polymers with different lengths $N$ are affected by active turbulent flows at varying extensile activity strengths $\zeta$.
We then explain the underlying mechanisms driving these conformational changes.
%%%%%%%%%%%%%%%%%%%%%%%%%%%%%%%%%%%%%%%%%%%%%%%%%%%%%%%%%%%%%
%% Conformation
%%%%%%%%%%%%%%%%%%%%%%%%%%%%%%%%%%%%%%%%%%%%%%%%%%%%%%%%%%%%%
\subsection{Activity-induced conformational changes}
The overall polymer extension is characterized by the average end-to-end distance $\re = \av{\abs{\vecP{\vec r}{N}-\vecP{\vec r}{1}}}$, where $\vecP{\vec r}{1}$ and $\vecP{\vec r}{N}$ are the positions of the first and the last monomers, respectively.
Here, $\av{\cdot}$ denotes the realization ensemble and temporal average.
Shorter polymers with $N\leq15$ monomers shrink relative to their thermal state with $\re/\reTrans<1$, where $\reTrans$ is the thermal end-to-end distance (\fig{fig:REE}a).
Long polymers ($N\geq100$) exhibit a predominantly extensional response.
Intermediate-length polymers ($15<N<100$) display a non-monotonic response: they initially contract but expand at higher activity.
The conformational response to activity depends strongly on polymer length: flexible polymers undergo a collapse–stretch transition.

This length-dependent behavior is both nontrivial and revealing.
It indicates that polymers do not respond to activity in a universal way, but instead exhibit distinct conformational regimes depending on their size.
While previous work has reported both swelling and compaction in active media, these effects were typically linked to particle density or activity strength, or focused on a specific chain length without identifying a systematic organizing principle~\cite{kaiser2014unusual,mousavi2021active,anderson2022polymer}.
Here, we observe a clear transition from collapse to extension as polymer length increases.
This indicates that activity strength alone does not determine the polymer's response; instead, the observed behavior points to an underlying mechanism involving a competition between length scales.

While the activity $\zeta$ sets the strength of the driving forces in the fluid, it does not capture how polymer size compares to the corresponding active length scale $\actlen$, which emerges only beyond the critical activity $\zeta^*$ in the turbulent regime (Appendix~\sctn{appendix:defects}).
In what follows, we focus on this regime.
It is the relative scale between the polymer size $\reTrans$ and the active length scale 
$\actlen$ that governs polymer conformational changes (\fig{fig:REE}b).
At small $\actlen/2\reTrans$, corresponding to long polymers compared to the active length scale, the chains are extended relative to their thermal size ($\re/\reTrans>1$).
As polymer length decreases relative to $\actlen$, the normalized end-to-end distance drops below one.
This indicates that polymers become more compact than the passive case when their size is smaller than the dominant flow structures.
Representative conformations illustrate this trend (\fig{fig:REE}b; inset).
Each polymer is visualized as a chain of monomers, overlaid with ellipses that represent passive and active conformations, constructed from the gyration tensor (\fig{fig:REE}c; Appendix \sctn{sec:Gyration}).
For the shorter polymer ($N=20$), the active conformation (\fig{fig:REE}b inset; pink ellipse) is visibly smaller than its passive counterpart (blue ellipse), reflecting its shrinkage in response to activity.
In contrast, for the longer polymer ($N=100$), the active ellipse is larger than the passive one, indicating that longer polymers are stretched by the active flows (\fig{fig:REE}b).
These comparisons highlight how the relative size of the polymer to the active flow structures governs whether activity leads to compaction or extension.

%%%%%%%%%%%%%%%%%%%%%%%%%%%%%%%%%%%%%%%%%%%%%%%%%%%%%%%%%%%%%
%%%%%%%%%%%%%%%%%%%%%%%%%%%%%%%%%%%%%%%%%%%%%%%%%%%%%%%%%%%%%
%% What is happening?
%%%%%%%%%%%%%%%%%%%%%%%%%%%%%%%%%%%%%%%%%%%%%%%%%%%%%%%%%%%%%
%%%%%%%%%%%%%%%%%%%%%%%%%%%%%%%%%%%%%%%%%%%%%%%%%%%%%%%%%%%%%
To examine local conformational correlations, we compute the discrete tangent–tangent correlation function \mbox{$C(n) = \av{\vecP{\vec t}{i} \cdot \vecP{\vec t}{i+n}}$}, which measures alignment between tangents $\vecP{\vec t}{i}$ and $\vecP{\vec t}{i+n}$ separated by $n$ bonds along the backbone.
For short polymers ($N=10$), the presence of active flows accelerates the decay of tangent–tangent correlation compared to the passive case (\fig{fig:correlations}a).
In contrast, for longer polymers ($N=100$) correlations persist over larger separations, and activity tends to enhance local alignment rather than disrupt it (\fig{fig:correlations}b).
The behavior of $C(n)$ thus provides insight into the contrasting conformational responses: in long polymers, enhanced long-range correlations account for their extended configurations.
Short polymers, by contrast, show a short-range decay in tangent–tangent correlations, reflecting their more contorted structures.
This decorrelation is most pronounced between adjacent bonds: the drop in 
$C(n=1)$ is significantly larger than the changes at longer separations.

To interpret this sharper decay, we consider the probability density of angles between consecutive bonds observed in short polymers (\fig{fig:correlations}d).
One possible mechanism for the reduced correlation is that activity increases the range of accessible bond angles.
In the passive case, the bond angle distribution is relatively flat, with most freely jointed angles being equally probable until contact between monomers becomes significant (\fig{fig:correlations}d; inset). 
Specifically, the excluded volume potential (Appendix \sctn{sec:MD}) sterically excludes angles greater than $2\pi/3$.
Activity does not reduce the excluded volume of the monomers; instead, the accessible bond-angle range remains mostly unchanged, while the likelihood of larger bending angles increases (\fig{fig:correlations}d).
The bond angle distribution for $N=10$ reveals only a slight broadening of the accessible angular range; however, it becomes substantially skewed toward larger angles, with a pronounced peak at $1.91 \pm 0.03$ radians, which is comparable to the contact angle $2\pi/3$.
This shift indicates that while extreme bond angles remain rare due to excluded volume, the probability of larger bending angles increases under activity, promoting the formation of folded conformations that contribute to overall compaction (\fig{fig:correlations}d; inset).

\begin{figure*}
    \centering
    \includegraphics[width=\textwidth]{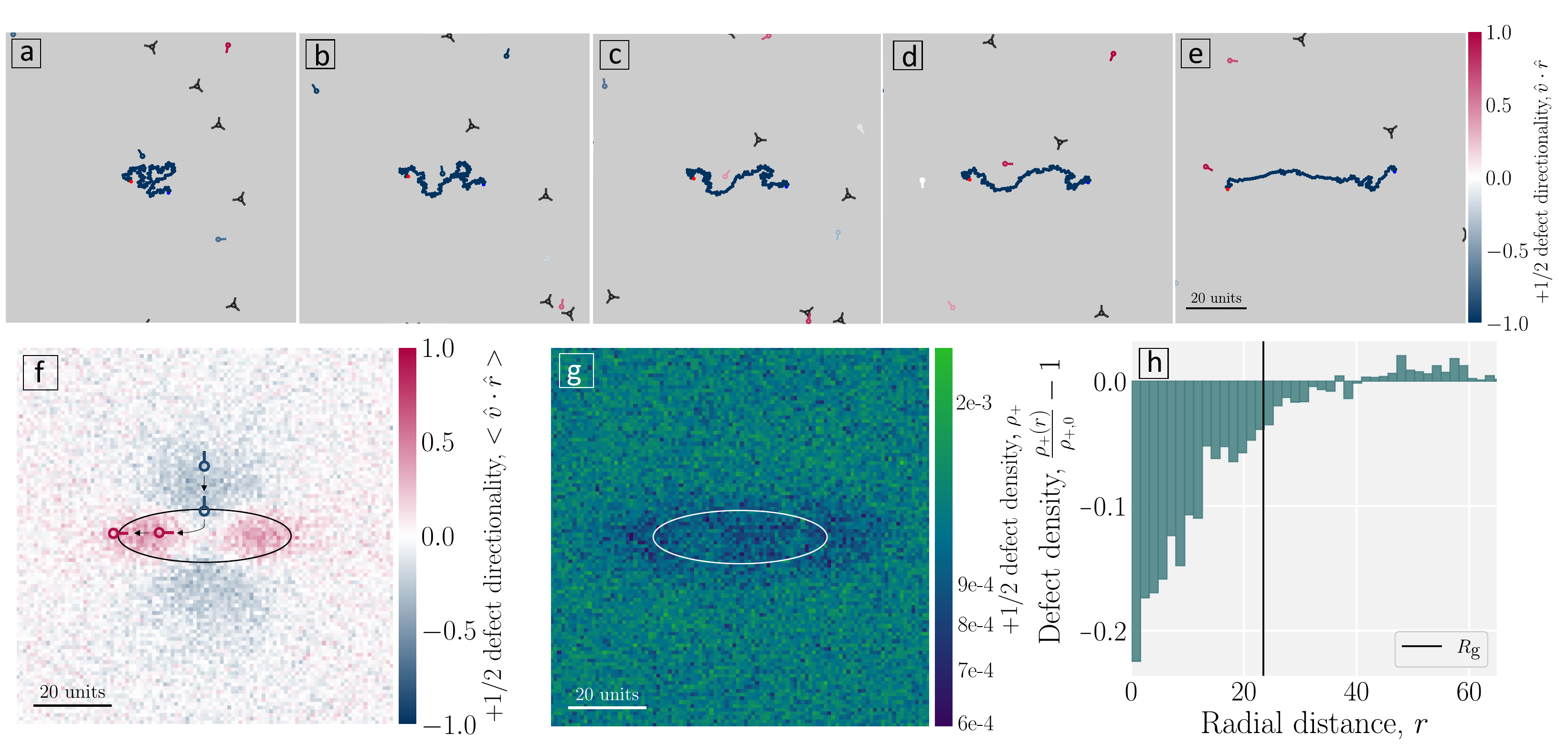}
	\caption{Spatial and angular organization of $\plus$ defects around long polymers.
    (a–e)
    Consecutive snapshots showing the positions and orientations of topological defects relative to a long polymer ($N=100$).
    To consistently analyze polymer–defect interactions, configurations are transformed into a reference frame where the polymer’s center of mass is placed at the origin and its major principal axis of the gyration tensor $\hat{e}_1$ (\fig{fig:REE}c) is aligned with the horizontal axis; defect positions and orientations are rotated accordingly.
    Plus-half defects are shown in color, coded by the alignment between their polarity vector $\hat{v}$ and the radial direction $\hat{r}$ pointing from the polymer center to the defect: red indicates outward motion ($\hat{v} \cdot \hat{r} > 0$), blue indicates inward orientation ($\hat{v} \cdot \hat{r} < 0$).
    Minus-half defects are in black for reference but are not color-coded.
    Snapshots are taken from simulations at $\zeta = 0.03$, with $20t_0$ between consecutive times.
    (f)
    Ensemble- and time-averaged spatial distribution of $\plus$ defect orientations relative to the polymer ($N=100$) for an activity of $\zeta = 0.08$.
    Configurations are transformed and defects are color-coded as in (a–e).
    Overlaid sketch shows typical inward (blue) and outward (red) $\plus$ defect motion.
    The ellipse outlines the average polymer extent, with semi-axes given by $\av{R_{\text{g},1}} + \sigma_{R_{\text{g},1}}$ and $\av{R_{\text{g},2}} + \sigma_{R_{\text{g},2}}$ along $\hat{e}_1$ and $\hat{e}_2$, respectively, where \( \sigma_{R_{\text{g},i}} \) denotes the standard deviation of 
    \( R_{\text{g},i} \) across time and ensemble realizations for \( i = 1, 2 \).
    (g)
    Ensemble- and time-averaged spatial distribution of $\plus$ defect positions relative to the polymer ($N=100$) for $\zeta = 0.08$.
    Configurations are transformed as in (a–e), and the ellipse is defined as in (f).
    (h) Radial density profile of $\plus$ defects $\rho_+(r)$, normalized by the far-field average defect density $\rho_{+,0}$ for a polymer with $N=100$ at $\zeta = 0.08$.
    The vertical line marks the polymer’s typical radius of gyration, $\av{R_{\text{g}}} + \sigma_{R_{\text{g}}}$.
    }
	\label{fig:LongPolymer}
\end{figure*}

The correlation for the next-nearest monomer $C(2)$ rises slightly above $C(1)$ (\fig{fig:correlations}a), suggesting the typical conformation is ``zig-zag-like''.
Indeed example trajectories have a zig-zag conformation (\fig{fig:correlations}c; $N=10$).
The curvature–curvature correlation \mbox{$C_k(n) =  \av{\vecP{\vec \kappa}{i} \cdot \vecP{\vec \kappa}{i+n}}$}, supports this interpretation quantitatively.
Here, the local curvature vector is $\vec{\kappa}_i = \vec{t}_{i+1} - \vec{t}_i$.
For short polymers $(N=10)$, activity increases the average curvature relative to the passive case, and the curvature–curvature correlation shows a negative peak at $n=1$ (\fig{fig:correlations}a; inset).
This anti-correlation indicates that consecutive bends tend to point in opposite directions, confirming the emergence of zig-zag motifs that drive chain compaction under activity.

Conversely, for the $N=100$ polymer, the probability of large bond angles diminishes with increasing activity (\fig{fig:correlations}e; inset), and the distribution becomes sharply peaked around zero, consistent with an extended, aligned configuration (\fig{fig:correlations}e, inset).
For long polymers, the rise in correlation of the next-nearest neighbors $C(2)$ above $C(1)$ is even more pronounced than for short polymers (\fig{fig:correlations}b vs. \ref{fig:correlations}a), which suggests that the conformation of the polymer is mostly long straight segments with sudden zig-zags at the shortest scales, and indeed this is consistent with qualitative snapshots (\fig{fig:correlations}c).
At $n=1$, the curvature–curvature correlation shows a mild anti-correlation, suggesting alternating bending directions (\fig{fig:correlations}b; inset).
However, the overall curvature magnitude is significantly lower than that of short polymers and that of the passive case, indicating that these zig-zags are shallow deflections rather than sharp folds.
Activity thus suppresses local bending in long chains, promoting extended configurations composed of mostly straight segments with occasional, gentle kinks--distinct from the tight bends observed in short polymers.
While the zig-zag conformation occurs because of steric interactions between segments, it is not responsible for the observed compaction, which persists even when local bond-level bending is suppressed (Appendix~\sctn{appendix:More-Stiff}).

%%%%%%%%%%%%%%%%%%%%%%%%%%%%%%
%%%%%%%   mechanism   %%%%%%%% 
%%%%%%%%%%%%%%%%%%%%%%%%%%%%%%
\subsection{Underlying mechanism}
%%%%%%%%%%%%%%%%%%%%%%%%%%%%%%
%%%%%%% defect features %%%%%% 
%%%%%%%%%%%%%%%%%%%%%%%%%%%%%%
Having characterized how polymer conformations respond to active flows, we now turn to understanding the underlying mechanisms driving these distinct behaviors.

%%%%%%%%%%%%%%%%%%%%%%%%%%%%%%
%%%%%%% Transition    %%%%%%%% 
%%%%%%%%%%%%%%%%%%%%%%%%%%%%%%
Given the unique ability of $\plus$ defects to self-propel and generate anisotropic, polar flow fields (Appendix \sctn{appendix:defects}), we focus our investigation on how they interact with polymers and contribute to their conformational dynamics in active turbulence.
For this purpose, we examine the spatial and orientational statistics of \( \plus \) defects relative to the polymer backbone.
%%%%%%%%%%%%%%%%%%%%%%%%%%%%%%
%%%%%%% Long polymers %%%%%%%% 
%%%%%%%%%%%%%%%%%%%%%%%%%%%%%%

Plus-half defects typically approach long polymers from the transverse sides, when the defect's polarity vector $\hat{v}$ is aligned with the polymer minor axis $\hat{e}_2$ (\fig{fig:LongPolymer}a-b; see Appendix~\sctn{appendix:defects} and \fig{fig:defectForces} for definition of $\hat{v}$).
The defect reorients upon reaching the chain (\fig{fig:LongPolymer}c) and proceeds to move along the backbone, closely following the polymer major axis $\hat{e}_1$ (\fig{fig:LongPolymer}d-e).
This directional interaction suppresses lateral undulations and progressively straightens the chain, contributing to an elongated configuration (\fig{fig:LongPolymer}d-e).
The polymer effectively guides the defect’s path, and in turn, the persistent active force of the defect irons out transverse fluctuations.

We quantify the orientation of each defect with respect to the polymer by calculating the dot product $\hat{v} \cdot \hat{r} $, where $\hat{r}$ is the radial unit vector pointing from the polymer center to the defect position.
This provides a scalar measure of directional alignment: values $\hat{v} \cdot \hat{r} > 0$ correspond to defects pointing away from the polymer (outward), while $\hat{v} \cdot \hat{r} < 0$ indicates inward orientations.   
Plus-half defects are more likely to point inward along polymer's minor axis \( \hat{e}_2 \) and outward along polymer's major axis \( \hat{e}_1 \) (\fig{fig:LongPolymer}f).
This anisotropic organization reflects a directional flow pattern—defects approach from the transverse sides and depart along the backbone-reinforcing stretching of the polymer.

Spatially, \( \plus \) defects are depleted within the polymer’s pervaded area (\fig{fig:LongPolymer}g). 
This depletion is visible in the time- and ensemble-averaged spatial defect density map $\rho_+(r)$.
The map shows a reduced presence of $\plus$ defects throughout the elliptical region defined by \( \langle R_{\text{g},1} \rangle + \sigma_{R_{\text{g},1}} \) and \( \langle R_{\text{g},2} \rangle + \sigma_{R_{\text{g},2}} \).
Here, \( R_{\text{g},i} \) are found from the eigenvalues of the gyration tensor and \( \sigma_{R_{\text{g},i}} \) are their standard deviation for \( i = 1, 2 \) (Appendix \sctn{sec:Gyration}).   
The radial defect density profile confirms this observation: the normalized density remains below the far-field average within a broad range around the polymer, including the region up to \( \langle R_{\text{g},1} \rangle + \sigma_{R_{\text{g},1}} \), marked by a vertical line (\fig{fig:LongPolymer}h).  
These results suggest that polymers effectively exclude \( \plus \) defects from their immediate surroundings.

%%%%%%%%%%%%%%%%%%%%%%%%%%%%%%
%%%%%%% Short polymers %%%%%%% 
%%%%%%%%%%%%%%%%%%%%%%%%%%%%%%
\begin{figure*}
    \centering
    \includegraphics[width=\textwidth]{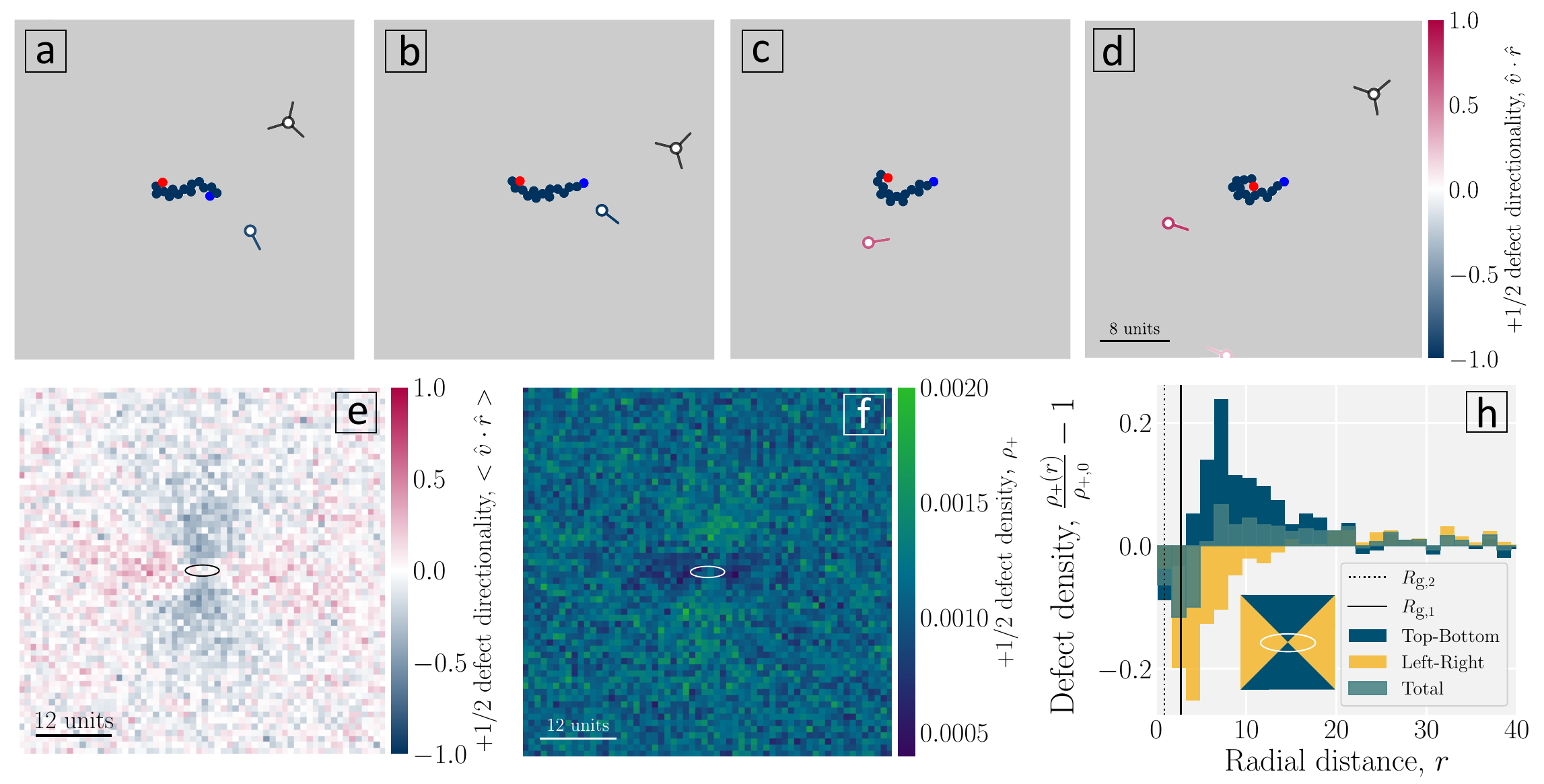}
	\caption{Spatial and angular organization of $\plus$ defects around short polymers.
    (a–d) Consecutive snapshots showing the positions and orientations of topological defects relative to a short polymer ($N=15$) at activity $\zeta = 0.03$.
     Configurations are transformed and defects are color-coded as in \fig{fig:LongPolymer}a-e.
    (e) Ensemble- and time-averaged spatial distribution of $\plus$ defect orientations relative to a polymer with $N=15$ monomers for activity $\zeta = 0.08$.
    The ellipse outlines the polymer’s pervaded area, using the same definition as in \fig{fig:LongPolymer}f.
    (f) Ensemble- and time-averaged spatial distribution of $\plus$ defect positions relative to polymer.
    (h) Radial distributions of \( \plus \) defect positions, resolved by their location around the polymer.  
    Defect are grouped into quadrants (Left, Right, Top, Bottom) based on their position; see inset for schematic.    
    Total presents the angle-averaged radial distribution, serving as a reference for isotropy.
    The dashed and solid vertical lines indicate the typical polymer extents along the minor and major axes, respectively:  
    \( \langle R_{\text{g},2} \rangle + \sigma_{R_{\text{g},2}} \) along \( \hat{e}_2 \) (dashed), and  
    \( \langle R_{\text{g},1} \rangle + \sigma_{R_{\text{g},1}} \) along \( \hat{e}_1 \) (solid), where \( \sigma_{R_{\text{g},i}} \) denotes the standard deviation of 
    \( R_{\text{g},i} \) across time and ensemble realizations for \( i = 1, 2 \).
    }
	\label{fig:ShortPolymer}
\end{figure*}

When a $\plus$ defect approaches a short polymer, the polymer chain lacks the longitudinal extent required for sustained alignment along the backbone.
Instead of traveling along the chain as seen in long polymers, the defect interacts only briefly with short polymers in a scattering-like event, inducing localized deformations (\fig{fig:ShortPolymer}a-d).
The polymer responds by curling into the surrounding flow field.
This interaction reflects the limited capacity of short polymers to guide or constrain defect motion.
The dynamics of defects around short polymers exhibit a weak, structured inward–outward circulation pattern (\fig{fig:ShortPolymer}e), similar to that observed around long polymers (\fig{fig:LongPolymer}f).
Defects tend to approach from the transverse sides (along $\hat{e}_2$) and depart along the backbone ($\hat{e}_1$).
The small size of the pervaded area limits the number of interacting defects, leaving too few data points to compute statistically robust measures of defect orientation.
However, visual inspection suggests a possible difference between short polymers (\fig{fig:ShortPolymer}e) and long polymers (\fig{fig:LongPolymer}f). For short polymers, outward-pointing defects often appear outside the polymer’s pervaded area, in contrast to long chains where they remain within it.
These observations suggest that the polymer’s short length prevents a defect from aligning with and traveling along its contour, as seen in long chains.
As a result, the mechanism observed in long polymers cannot operate here.
Instead, the interaction remains brief and localized, insufficient to enforce longitudinal stretching or suppress transverse fluctuations in the same manner as in long polymers.

On the other hand, the spatial distribution of $\plus$ defects around short polymers reveals a different pattern compared to long polymers.
The distribution shows subtle anisotropy (\fig{fig:ShortPolymer}f).
In contrast to the long-polymer case--where defects are excluded both from the interior and a surrounding shell (\fig{fig:LongPolymer}g)--the short-polymer regime exhibits no such symmetric depletion.
Instead, defect density is weakly enhanced along the polymer’s minor axis $\hat{e}_2$, while the central region remains relatively depleted, consistent with local spatial exclusion by the polymer backbone.
To quantify this anisotropy, we resolve the radial density of $\plus$ defects according to their position relative to the polymer principal axes (\fig{fig:ShortPolymer}h).
Specifically, the domain is divided into quadrants—Top, Bottom, Left, and Right— relative to the polymer’s orientation (\fig{fig:ShortPolymer}h; inset), and the \mbox{left–right} and \mbox{top–bottom} pairs are combined due to symmetry.
This allows us to assess whether defects preferentially accumulate along the polymer longitudinal (major axis $\hat{e}_1$) or transverse (minor axis $\hat{e}_2$) directions.
The resulting profiles reveal a strong directional dependence: defects are depleted along the polymer major axis but show enhanced density just beyond the extent of the minor axis.
The vertical dashed and solid lines mark the characteristic spatial extents along the minor and major axes, respectively.
The top–bottom profile peaks just beyond the minor-axis extent, highlighting the spatial offset of this anisotropic enrichment.
These features confirm that defects tend to avoid the polymer backbone while accumulating near its transverse sides.
The total angle-averaged radial profile shows both depletion within and just outside the polymer, along with a weak shoulder at intermediate distances, indicating that anisotropic structure persists even after angular averaging, albeit more subtly than in the quadrant-resolved data (\fig{fig:ShortPolymer}h).

This anisotropic enrichment of $\plus$ defects near the polymer’s transverse sides, coupled with their depletion along the longitudinal axis, reveals a directional imbalance in active stresses—suggesting an effective osmotic pressure directed inward along the minor axis.
Such directional crowding bends the chain inward and contribute to conformational shrinkage in short polymers, especially in the absence of longitudinal defect motion to counteract it.

%%%%%%%%%%%%%%%%%%%%%%%%%%%%%%%%%%%%%%%%%%%%%%%%%%%%%%%%%%%%%
%%%%%%%%%%%%%%%%%%%%%%%%%%%%%%%%%%%%%%%%%%%%%%%%%%%%%%%%%%%%%
%% Conclusion
%%%%%%%%%%%%%%%%%%%%%%%%%%%%%%%%%%%%%%%%%%%%%%%%%%%%%%%%%%%%%
%%%%%%%%%%%%%%%%%%%%%%%%%%%%%%%%%%%%%%%%%%%%%%%%%%%%%%%%%%%%%
\section*{Conclusion}
This study explored the conformations of flexible polymers suspended in an active nematic liquid crystal.
Activity induces a length-dependent coil–stretch transition: short polymers shrink, long polymers extend, and intermediate chains exhibit non-monotonic behavior as activity increases.
These transitions are controlled by the ratio of polymer size $\reTrans$ to the active length scale $\actlen$: polymers larger than the characteristic flow structures ($\actlen/2\reTrans<1$) are stretched, while smaller ones ($\actlen/2\reTrans>1$) shrink.
% These transitions are controlled by the relative scale of the polymer size $\reTrans$ and the characteristic active length of the flow $\actlen$.
We observe that short polymers exhibit enhanced bending and rapid angular decorrelation, which we refer to as contortions, while long polymers maintain persistent alignment along their contour.
Intermediate chains transition between these regimes, with activity shifting their structure from compact to elongated.

This behavior is mediated by interactions with $\plus$ defects that structure the active flow.
As \( \plus \) defects traverse long polymers, their persistent propulsion and polar flow fields impose shear-like deformations that stretches the polymer along its backbone.
By ironing out transverse undulations and promoting alignment with the flow, these directional interactions give rise to sustained extensional configurations.
In contrast, short polymers experience only brief, asymmetric interactions: defects accumulate near their sides but do not generate sustained alignment, leading instead to compaction.  
These findings identify a defect-driven mechanism for conformational control, governed by the interplay between polymer size and flow coherence. This defect-mediated mechanism of polymer remodeling relies on action at a distance from the flow field, and is therefore fundamentally distinct from the one leading to hairpin formation which was identified in polymers activated by molecular motors~\cite{foglino2019nonequilibrium}, where hairpins arise from tangential forces exerted by the polymer-bound motors on the substrate as they move.
Whilst both mechanisms remodel polymer conformations, motor-mediated activity generates hairpins that soften their polymeric substrate, whereas defect-mediated contortions can either stiffen or soften the polymer depending on its size.

This work therefore highlights a unifying mechanism for polymer conformational control in active matter: the competition between internal polymer length scales and the emergent structural scale of the active medium.  
This perspective helps reconcile the diverse behaviors reported in both dry and wet active systems—ranging from compaction to swelling—by framing them in terms of length-scale competition rather than system-specific parameters~\cite{kaiser2014unusual,harder2014activity,mousavi2021active,yadav2024passive,liu2019configuration}. 

Future studies could explore whether similar transitions arise in polymers with intrinsic stiffness, active propulsion, or other internal structure, and how these effects are modified by confinement or dimensionality~\cite{miranda2023self,mori2023viscoelastic,martin2024tangentially}. 
More generally, tuning activity to modulate polymer conformation may enable new strategies for the design of self-organizing or responsive materials~\cite{kumar2024emergent}, as well as improved understanding of non-equilibrium steady-state conformations in active biological environments~\cite{gompper20202020}.

%%%%%%%%%%%%%%%%%%%%%%%%%%%%%%%%%%%%%%%%%%%%%%%%%%%%%%%%%%%%%
%%%%%%%%%%%%%%%%%%%%%%%%%%%%%%%%%%%%%%%%%%%%%%%%%%%%%%%%%%%%%
%% Acknowledgments
%%%%%%%%%%%%%%%%%%%%%%%%%%%%%%%%%%%%%%%%%%%%%%%%%%%%%%%%%%%%%
%%%%%%%%%%%%%%%%%%%%%%%%%%%%%%%%%%%%%%%%%%%%%%%%%%%%%%%%%%%%%
\section*{Acknowledgments}
This research has received funding from the European Research Council (ERC) under the European Union’s Horizon 2020 research and innovation program (Grant agreement No. 851196).
We acknowledge useful discussions with Benjam\'{i}n
Loewe, Oleksandr Baziei, Manasa Kandula, Patrick Warren and Wilson Poon.
For the purpose of open access, the author has applied a Creative Commons Attribution (CCBY) license to any Author Accepted Manuscript version arising from this submission.

\section*{Conflicts of interest}
There are no conflicts to declare. 
%%%%%%%%%%%%%%%%%%%%%%%%%%%%%%%%%%%%%%%%%%%%%%%%%%%%%%%%%%%%%
%%%%%%%%%%%%%%%%%%%%%%%%%%%%%%%%%%%%%%%%%%%%%%%%%%%%%%%%%%%%%
%%%%%%%%%%%%%%%%%%%%%%%%%%%%%%%%%%%%%%%%%%%%%%%%%%%%%%%%%%%%%
%% Appendix
%%%%%%%%%%%%%%%%%%%%%%%%%%%%%%%%%%%%%%%%%%%%%%%%%%%%%%%%%%%%%
%%%%%%%%%%%%%%%%%%%%%%%%%%%%%%%%%%%%%%%%%%%%%%%%%%%%%%%%%%%%%
%%%%%%%%%%%%%%%%%%%%%%%%%%%%%%%%%%%%%%%%%%%%%%%%%%%%%%%%%%%%%
\section*{Methods}
\appendix
%%%%%%%%%%%%%%%%%%%%%%%%%%%%%%%%%%%%%%%%%%%%%%%%%%%%%%%%%%%%%
\section*{Active Nematic Multi-Particle Collision Dynamics (AN-MPCD)}
\label{sec:AN-MPCD}
%%%%%%%%%%%%%%%%%%%%%%%%%%%%%%%%%%%%%%%%%%%%%%%%%%%%%%%%%%%%%
This method represents the fluid as point particles, each defined by its mass $m_i$, position $\vec{r_i}$, velocity $\vec{v}_i$, and nematic orientation $\vec{u}_i$~\cite{shendruk2015multi}.
The number of lines beneath a symbol indicates the tensor rank. While time is discretized into finite intervals $\Delta t$, all other quantities evolve continuously.
The time evolution of the system follows a two-step process: a streaming step and a collision step.
In the steaming step, particle positions evolve ballistically as
\begin{align}
    \vecP{\vec r}{i}\Par{t+\Delta t} &= \vecP{\vec r}{i}\Par{t} + \vecP{\vec v}{i}\Par{t} \Delta t.   
\end{align}
During the collision step, coarse-grained interactions between particles redistribute both momentum and orientation.
Particles are binned into cells (labeled $c$) of size $a$ containing $N_c(t)$ particles at a given time $t$.
A random grid shift is applied to maintain Galilean invariance~\cite{ihle2001stochastic}.
Interactions occur exclusively among particles within the same cell, ensuring that all particles in a given cell participate in the interaction process.

The collision step involves two phases: $(1)$ momentum exchange and $(2)$ orientation fluctuations.
\paragraph*{(1) Momentum exchange.} The velocity of particle $i$ within cell $c$ is updated as $\vec{v}_i(t+\Delta t) = \vec{v}^{\text{cm}}_c(t)+ \vec{\VCO}_{i,c}$, where $\vec{v}^{\text{cm}}_c(t) = \av{\vec{v}_i(t)}_c$ is the velocity of the cell center of mass at time $t$, $\av{\cdot}_c$ represents the mass-weighted average over the $N_c$ particles instantaneously within the cell $c$ and $\vec{\VCO}_{i,c}$ is the collision operator.
The collision operator consists of a passive contribution ${\vec{\VCO}}^{\text{P}}_{i,c}$ and an active part ${\vec{\VCO}}^{\text{A}}_{i,c}$.
We assume all the fluid particles have the same mass $m$. 
For the passive contribution, a modified angular-momentum conserving Andersen-thermostatted collision operator is employed. 
In the absence of angular momentum conservation, the Andersen-thermostatted collision operator is ${\vec{\VCO}}^{\text{P}}_{i,c} = {\vec{v}^{\text{ran}}_i} - \av{{\vec{v}^{\text{ran}}_i}}_c$, where the components of ${\vec{v}^{\text{ran}}_i}$ are Gaussian random numbers 
with variance $\kbt/m$ and zero mean~\cite{Noguchi2007,allahyarov2002mesoscopic}. 
Although the individual velocity of each particle is randomized, subtracting $\av{{\vec{v}^{\text{ran}}_i}}_c$ assures that the cell center-of-mass velocity---and consequently the total momentum---remains unchanged~\cite{Noguchi2007}. 
To conserve angular momentum, an additional term must be added to the collision operator to remove any angular velocity generated by the collision~\cite{gotze2007relevance}. 
In an isotropic system, the change in the angular momentum can rise due to the stochastically generated velocities 
$ \delta \vec{\mathcal{L}}_{\text{vel}} = m \sum_{j=1}^{N_c(t)} \vec{r}_{j,c} \times (\vec{v}_j - {\vec{v}^{\text{ran}}_j})$, where $\vec{r}_{i,c} = \vec{r}_i - \av{\vec{r}_i}_c$ is the relative position of particle $i$ with respect to the center of mass of all particles in the cell. 
In a nematic system, changes in the orientation will contribute additional angular momentum $\delta\vec{\mathcal{L}}_{\text{ori}}$, which will be elaborated below. 
For angular-momentum conserving Andersen-thermostatted AN-MPCD the passive collision operator is
\begin{align}
    \label{eq:CollisionOperator}
    {\vec{\VCO}}^{\text{P}}_{i,c} &= {\vec{v}^{\text{ran}}_i} - \av{{\vec{v}^{\text{ran}}_i}}_c + \Par{{\tens{I}^{-1}_c} \cdot \Par{\delta \vec{\mathcal{L}}_{\text{vel}}+\delta\vec{\mathcal{L}}_{\text{ori}}}} \times \vec{r}_{i,c},
\end{align} 
where the moment of inertia tensor $\tens{I}_c$ is for the particles in cell $c$. 

Activity is introduced as a force dipole that locally injects energy at the scale of one MPCD cell without affecting momentum, applied by the active component of the collision operator
\begin{align}
    \label{eq:ActiveNematic}
    {\vec{\VCO}}^{\text{A}}_{i,c} &= 
    \zeta_c \Delta t \Par{
    \frac{\kappa_i}{m_i}-\frac{\av{\kappa_i}_c}{\av{m_i}_c}
    }
    \vecP{\vec{n}}{c},
\end{align}
where $\zeta_c$ sets the strength of cellular force dipole ~\cite{kozhukhov2022mesoscopic}.
To determine the direction of the active force acting on particle $i$, cell $C$ is divided by a plane passing through its center of mass, with a normal vector $\vecP{\vec{n}}{c}$. 
Depending on whether $\vecP{\vec{r}}{i}$ lies above or below this plane, the value of $\kappa_i(\vecP{\vec{r}}{i}, \vecP{\vec{n}}{c})$ is assigned as $\pm 1$, indicating the direction of the active force.
The first term in \eq{eq:ActiveNematic} represents individual impulses per unit mass on particle $i$, and the second term removes any residual impulse to locally conserve momentum. 

The strength of the active dipole in cell $c$ can be assigned in various ways, depending on the physics of interest.
For instance, if all $N_c$ fluid particles within a cell are treated as active units, $\zeta_c = \sum_{j=1}^{N_c} \zeta_j$.
In this scheme there is a positive feedback between activity and density that can exacerbate density inhomogeneities\cite{kozhukhov2024mitigating}.
Since, density inhomogeneities affects the dynamics of inclusions~\cite{zantop2021multi}, we have chosen a modulated strength to mitigate activity-induced density fluctuations~\cite{kozhukhov2024mitigating}.
This modulated activity is given by
\begin{equation}
    \zeta_c = S_c (N_c; \sigma_p,\sigma_w) \sum_{j=1}^{N_c} \zeta_j,
\end{equation}
where $S_c$ is a sigmoidal function expressed as
\begin{equation}
    S_c (N_c; \sigma_p,\sigma_w) = \frac{1}{2}
    \Par{
    1 - \tanh
    \Par{
    \frac{N_c - \av{N_c} \Par{1+\sigma_p}}{\av{N_c}\sigma_w}
    }}.
\end{equation}
This function compares a cell instantaneous number density $N_c(t)$ to the system-wide average $\av{N_c}$ and returns a value between 0 and 1.
The parameters $\sigma_p$ and $\sigma_w$ control the position and width of the sigmoid drop. 
$\sigma_p$ sets the midpoint: $\sigma_p=0$ places it at $\av{N_c}$, while $\sigma_p>0$ ($\sigma_p<0$) shifts it to higher (lower) densities.

\paragraph*{(2) Orientation fluctuations.} Orientation of particle $i$ updates as $\vecP{\vec u}{i}\Par{t+\Delta t} = \OCO$, where $\OCO$ is the nematic collision operator acting on particles in cell $c$.
The collision operator acts as a rotation, modifying the orientation of each AN-MPCD particle during the time step $\Delta t$.
The reorientation process involves (i) a \textit{stochastic} contribution $\Par{\delta \vecP{\vec u}{i}^{ST}/\delta t}$ and (ii) a \textit{flow-induced} contribution $\Par{\delta \vecP{\vec u}{i}^{J}/\delta t}$.

The \textit{stochastic} orientations are drawn from the canonical distribution of the Maier-Saupe mean-field approximation $f_{\text{ori}}\left(\vec{u}_i\right) = f_0 \exp \Par{Us_c \Par{\vecP{\vec u}{i} \cdot \vecP{\vec n}{c}}^2/\kbt}$, which is centered around $\vecP{\vec n}{c}$. 
Here, $f_0$ is the normalization constant and $U$ is the mean-field interaction constant that governs the alignment strength between the MPCD nematogens~\cite{shendruk2015multi}.
The local director $\vecP{\vec n}{c}$ is the eigenvector of the cell's tensor order parameter $\tens{Q}_c\Par{t} = \frac{1}{d-1} \av{d\vecP{\vec u}{i}\Par{t}\vecP{\vec u}{i}\Par{t} - \tens{1}}_{c}$, where $\tens{1}$ is the identity tensor in $d$ dimensions.
This eigenvector is associated with the largest eigenvalue $s_c$, which is the scalar order parameter.

The \textit{flow-induced} reorientations arise from the response of MPCD nematogens to velocity gradients. 
These orientation changes are captured through Jeffrey's equation 
\begin{align}
    \frac{\delta {\vecP{\vec u}{i}}^J}{\delta t} &= \alpha \left[ \vecP{\vec u}{i} \cdot \tens{\Omega}_c + \lambda \Par{\vecP{\vec u}{i} \cdot \tens{E}_c - \vecP{u}{i}\vecP{u}{i}\vecP{u}{i}\colon \tens{E}_c} \right] , 
\end{align} 
for a bare tumbling parameter $\lambda$ and shear susceptibility $\alpha$ in a flow with strain rate tensor $\tens{E}_c\Par{t} = \Par{\nabla \vecP{\vec v}{c} + \Par{\nabla \vecP{\vec v}{c}}^T}/2$ and rotation rate tensor  $\tens{\Omega}_c\Par{t} = \Par{\nabla \vecP{\vec v}{c} - \Par{\nabla \vecP{\vec v}{c}}^T}/2$. 
The rotations of nematogens generate hydrodynamic motion, known as backflow, which is accounted for by the change in angular momentum 
$\delta\vec{\mathcal{L}}_{\text{ori}} = - \delta t \Par{\gamma_R \sum_{j=1}^{N_c(t)} \vecP{u}{i}\Par{t} \times \vecP{\vec {\dot{u}}}{i}}$, where $\gamma_R$ is the rotational friction coefficient and $\vecP{\vec {\dot{u}}}{i} = \Par{\delta \vecP{\vec u}{i}^{ST}/\delta t} + \Par{\delta \vecP{\vec u}{i}^{J}/\delta t}$~\cite{shendruk2015multi,armendariz2021nonequilibrium,Noguchi2007}.

The AN-MPCD method allows for simulations of active nematic fluids, exhibiting the key characteristics of active nematic turbulence~\cite{kozhukhov2022mesoscopic}.
The MPCD cell size $a$ sets the unit of length, thermal energy $\kbt$ defines the energy unit and the mass unit is fluid particle mass $m$.
These units determine other simulation units, including units of time $t_0 = a\sqrt{m/\kbt}$.
The time step is set to $\Delta t = 0.1 t_0$.
The rotational friction coefficient is $\gamma_\text{R} = 0.01\kbt t_0$, chosen small to minimize backflow effects.
The bare tumbling parameter is $\lambda = 2$ and the shear susceptibility is $\alpha = 0.5$.
The average number of AN-MPCD particles per cell is $\av{N_c} = 20$, where $\av{\cdot}$ denotes the average over the entire system. The nematic interaction constant $U = 10\kbt$ is deep within the nematic phase~\cite{shendruk2015multi}.
We explore a range of extensile activities $\zeta \in [0, 0.08]$, with $\sigma_p = 0.5 $ and $\sigma_w = 0.1$ in activity modulation function. 
These values are shown to give optimal density behavior without lowering effective activity~\cite{kozhukhov2024mitigating}.

%%%%%%%%%%%%%%%%%%%%%%%%%%%%%%%%%%%%%%%%%%%%%%%%%%%%%%%%%%
%% Polymer model
%%%%%%%%%%%%%%%%%%%%%%%%%%%%%%%%%%%%%%%%%%%%%%%%%%%%%%%%%%
\section*{Polymer model}
\label{sec:MD}

Each monomer $j$ of mass $M$ obeys a Langevin-like equation of motion
\begin{align}
    \label{eq:EquationOfMotion}
    M{\vecP{\vec{\Ddot{r}}}{j}} = - \vec{\nabla} V_j+
    - \gamma_\text{eff}
    \Par{ \vecP{\vec{\dot{r}}}{j} - \vec{v}^{\text{cm}}_c(t) } + \eta_j(t),
\end{align}
where \( V_j \) is the total potential acting on monomer \( j \).
The second term describes viscous drag with an effective friction coefficient \( \gamma_\text{eff} \), defined relative to the center-of-mass velocity of the MPCD cell \( c \) containing the monomer, \( \vec{v}^{\text{cm}}_c(t) \). 
The last term \( \vec{\eta}_j(t) \), is a stochastic force that captures momentum exchange with the solvent and ensures thermal fluctuations consistent with the MPCD thermostat. 
Both \( \gamma_\text{eff} \) and the noise statistics are determined by the specific MPCD collision rules.
\EQ{eq:EquationOfMotion} offers a coarse-grained description of polymer dynamics in a thermally fluctuating solvent, where hydrodynamic and thermal effects are encoded in the dissipative and stochastic forces.
The beads are connected by a finitely extensible nonlinear elastic (FENE) bond potential~\cite{Grest1986,Kremer1990,Slater2009}
\begin{align}
    \label{eq:FENE}
    V_{\text{FENE}}(r_{jl}) = -\frac{k_\text{FENE}}{2} {r_0}^2 \ln\left(1 - \frac{r_{jl}^2}{{r_0}^2}\right),
\end{align}
where $k_\text{FENE}$ is the bond strength, $r_0$ the maximum bond length and $r_{jl} = \abs{ \vec{r}_j - \vec{r}_l }$ the distance between monomers $j$ and $l$, with $l = j - 1$ for bonded pairs.
Excluded volume effects are modeled using a purely repulsive variant of the Lennard-Jones potential, known as the Weeks-Chandler-Andersen (WCA) potential~\cite{royall2024colloidal,Weeks1971}
\begin{equation}
    \label{eq:WCA}
    V_{\text{WCA}}(r_{jl}) = 4\epsilon 
    \begin{cases}
          \left(\frac{\sigma}{r_{jl}}\right)^{12} - \left(\frac{\sigma}{r_{jl}}\right)^6 + \frac{1}{4}, & r_{jl}<\sigma_{\text{co}}\\
         0, & r_{jl} \geq \sigma_{\text{co}},
    \end{cases}
\end{equation}
where \(\epsilon\) controls the strength of the repulsive potential and \(\sigma\) represents the effective size of a bead.
The potential smoothly approaches zero when the distance between beads $r_{jl}$ exceeds the cutoff distance $\sigma_{\text{co}} = 2^{1/6}\sigma$.

In our simulations, the MD bead size is $\sigma  = 1a$ and the repulsive strength is $\epsilon = 1\kbt$.
The maximum bond length is $r_0 =1.5a$, and the bond strength is $k_\text{FENE}=120\kbt/a^2$.
Monomer mass is $M =10m$.
The degree of polymerization is $N={5,10,15,20,25,50,100,200}$, corresponding to a contour length $ = (N-1) b$, where $b$ is the average bond length.
The MD time step is ${\Delta t}_{\text{MD}} = 0.002t_0$, requiring $50$ MD iterations per AN-MPCD iteration. 

%%%%%%%%%%%%%%%%%%%%%%%%%%%%%%%%%%%%%%%%%%%%%
%%%% Setup
%%%%%%%%%%%%%%%%%%%%%%%%%%%%%%%%%%%%%%%%%%%%%
\section*{System setup}
\label{sctn:setup}

Polymers are embedded in a periodic box of size $\SystemSize\times \SystemSize = 220 a \times 220 a$ with periodic boundary conditions, unless stated otherwise.
The fluid particles are initialized with random positions, Maxwell-Boltzmann distributed velocities, and uniform orientations along the $\hat{x}$-axis.
The polymer is initialized in a fully extended conformation, aligned with the global nematic direction, centered at the simulation box's center, and allowed to relax.
Data is recorded once the system reaches its steady state, which is identified via an iterative procedure.
In each iteration, the ensemble average is compared to the overall average from a candidate start time, across all repetitions.
The earliest time at which the ensemble average drops below the overall average is selected as a new candidate start time.
This procedure is repeated, updating the reference time in each cycle, until it converges.
The final reference time marks the onset of steady state.
Forty repeats for each set of parameters are simulated, each lasting $2.2-2.5\times 10^4 t_0$.
All simulations are performed using a custom-developed AN-MPCD/MD solver~\cite{shendrukLabGithub}.

%%%%%%%%%%%%%%%%%%%%%%%%%%%%%%%%%%%%%%%%%%%%%%%%%%%%%%%%%%%%%
\section*{Polymer shape analysis from the gyration tensor}
\label{sec:Gyration}
%%%%%%%%%%%%%%%%%%%%%%%%%%%%%%%%%%%%%%%%%%%%%%%%%%%%%%%%%%%%%
Polymer conformations are represented by ellipses constructed from the eigenvalues and eigenvectors of the time-averaged two-dimensional gyration tensor \( \tens{G} \). 
This tensor is defined as
\begin{equation}
    \tens{G} = \frac{1}{N} \sum_{j=1}^N (\vec{r}_j - \vec{r}_\text{cm})(\vec{r}_j - \vec{r}_\text{cm})^\top,
\end{equation}
where \( \vec{r}_j \) is the position of monomer \( j \), and \( \vec{r}_\text{cm} = {\av{\vec{r}_j}}_N \) denotes the polymer's center of mass.
The eigenvalues \( \lambda_1 \geq \lambda_2 \) and corresponding orthonormal eigenvectors \( \hat{e}_1 \) and \( \hat{e}_2 \) of \( \tens{G} \) define the principal directions and shape of the polymer.
The root mean square extensions along these directions are given by
\begin{equation}
    R_{\text{g},1} = \sqrt{\lambda_1}, \qquad R_{\text{g},2} = \sqrt{\lambda_2},
\end{equation}
which characterize the polymer’s spatial extent along \( \hat{e}_1 \) and \( \hat{e}_2 \), respectively (\fig{fig:REE}c).
The scalar radius of gyration, which quantifies the overall spatial extent of the polymer independent of orientation, is given by the trace of the gyration tensor
\begin{equation}
    R_\text{g}^2 = \text{Tr}(\tens{G}) = \lambda_1 + \lambda_2.
\end{equation}
To account for conformational fluctuations, we compute the standard deviations \( \sigma_{R_{\text{g},1}} \) and \( \sigma_{R_{\text{g},2}} \) of these extensions across time and ensemble realizations. 
The resulting ellipse illustrates the polymer’s average anisotropic shape, with semi-axes \( \langle R_{\text{g},1} \rangle + \sigma_{R_{\text{g},1}} \) and \( \langle R_{\text{g},2} \rangle + \sigma_{R_{\text{g},2}} \) aligned along \( \hat{e}_1 \) and \( \hat{e}_2 \).
These principal components characterize both the elongation and overall orientation of the polymer.

%%%%%%%%%%%%%%%%%%%%%%%%%%%%%%%%%%%%%%%%%%%%%%%%%%%%%%%%%%%%%
\section*{Defects and length scale in active turbulence}
\label{appendix:defects}
%%%%%%%%%%%%%%%%%%%%%%%%%%%%%%%%%%%%%%%%%%%%%%%%%%%%%%%%%%%%%
\begin{figure}
    \centering
    \includegraphics[width=\linewidth]{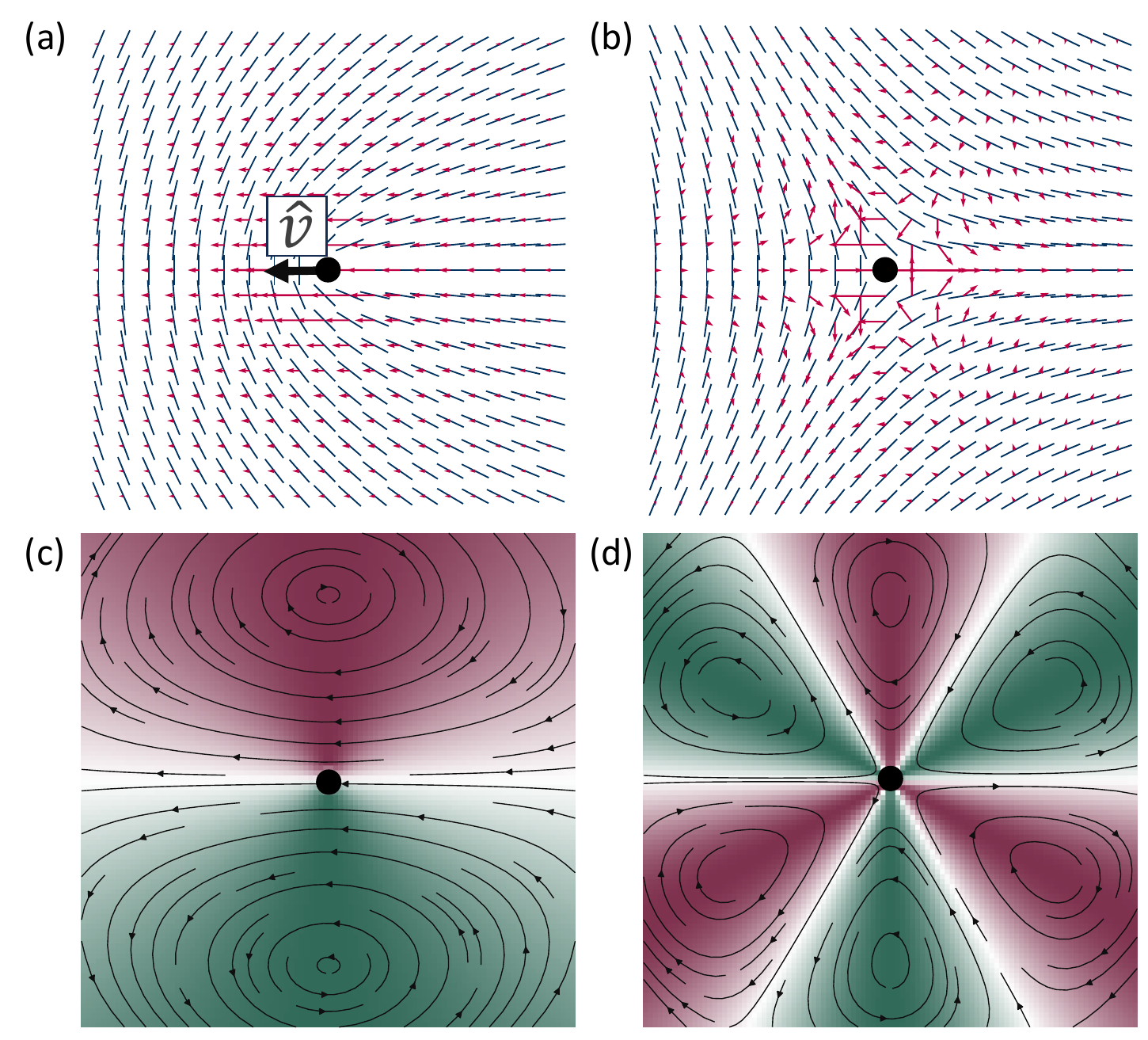}
    \caption{Activity-generated force and flow fields around isolated $+1/2$ and $-1/2$ defects. 
    (a,b) Active force fields (red arrows) for (a) $+1/2$ and (b) $-1/2$ defects. 
    The director field is indicated by blue lines, and the defect cores by circles. 
    The black arrow in panel~(a) marks the polarity vector $\hat{v}$ of the $+1/2$ defect. 
    (c,d) Active flow fields (arrows) for (c) $+1/2$ and (d) $-1/2$ defects, with background color denoting vorticity.}
    \label{fig:defectForces}
\end{figure}

Active turbulence is sustained by the continual creation and annihilation of topological defects in the nematic director field~\cite{thampi2014instabilities}.  
These defects are singularities where the director becomes ill-defined and the scalar order parameter \( s \) drops sharply, approaching zero.
In two dimensions, defects are point-like and carry quantized charges, typically \( \pm 1/2 \).  
They organize local flow fields~\cite{giomi2014defect}, mediate momentum transfer, and drive the spatiotemporal dynamics of active turbulence~\cite{serra2023defect,tan2019topological}.

The active force density generated by topological defects originates from gradients in the nematic stress, which in turn depend on director distortions near the defect core~\cite{giomi2014defect}.
The dynamics of active defects depend critically on their topological charge~\cite{head2024spontaneous}.
For \( +1/2 \) defects, the force distribution is polar and directed along the defect’s symmetry axis, which defines the polarity vector $\hat{v}$ (\fig{fig:defectForces}a, black arrow). 
This asymmetry results in self-propulsion, with the defect moving persistently along its head-tail axis.  
The associated velocity field exhibits dipolar symmetry, with a pair of counter-rotating vortices centered around the defect (\fig{fig:defectForces}c).  
Due to the logarithmic nature of 2D hydrodynamic interactions, the flow velocity at the core scales with system size $\vec{v}_+ = \zeta \SystemSize/\Par{4\eta} \hat{v}$.
In contrast, \( -1/2 \) defects have a threefold symmetric force field and do not self-propel (\fig{fig:defectForces}b).  
Their velocity field has a six-vortex pattern (\fig{fig:defectForces}d), resulting in zero net motion at the core $\vec{v}_- = \vec{0}$.
Any motion of the defect core arises solely from advection by surrounding flows or elastic interactions with other defects~\cite{giomi2014defect}.
In both $\pm1/2$ cases, the force magnitude is strongest at the defect core and decays with $1/r$.
The polarity of the \( +1/2 \) defect and the directional flow it generates are essential to understanding its role in driving polymer alignment and extension.

Despite the apparent disorder of active nematic turbulence, there exists a distinctive spatial coherence~\cite{doostmohammadi2018active,sanchez2012spontaneous,henkin2014tunable}.
The competition between activity ($\sim |\zeta|$), which drives flows by distorting the director field, and nematic elasticity $(\sim K/\actlen^2)$, which resists such deformations, gives rise to an active length scale~\cite{hemingway2016correlation,ramaswamy2010mechanics,thampi2014instabilities,thampi2014vorticity}
\begin{equation}
    \label{eq:activeLength}
    \actlen = \sqrt{\frac{K}{|\zeta|}}, 
\end{equation}
which governs the spatial structure of flows in the turbulent regime~\cite{doostmohammadi2018active}.
In simulations, we estimate \( \actlen \) from the steady-state defect density using \( \actlen = \sqrt{A/\av{N_\text{d}}} \), where \( A = \SystemSize \times \SystemSize \) is the system area and \( \av{N_\text{d}} \) is the time-averaged number of topological defects.
As activity decreases, \( \actlen\) grows and eventually exceeds the system size \( \SystemSize \), at which point turbulent dynamics vanish and residual flows are governed by finite-size effects~\cite{alert2022active}.  
% are driven by finite-size confinement and periodic boundary conditions
For the system studied here, active turbulence emerges above a critical activity of \( \zeta^* = 0.03 \) (\fig{fig:REE}a, dashed line).
   
%%%%%%%%%%%%%%%%%%%%%%%%%%%%%%%%%%%%%%%%%%%%%%%%%%%%%%%%%%%%%
\section*{Probability densities of polymer end-to-end distance}
\label{appendix:pdfs}
%%%%%%%%%%%%%%%%%%%%%%%%%%%%%%%%%%%%%%%%%%%%%%%%%%%%%%%%%%%%%
\begin{figure}
    \centering
    \includegraphics[width=0.8\linewidth]{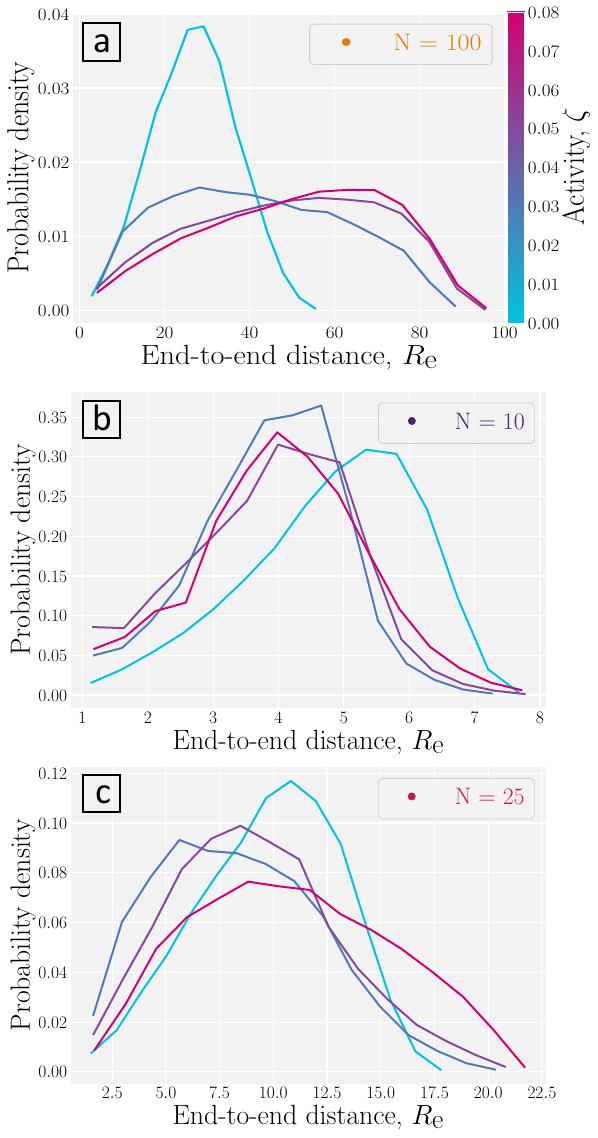}
    \caption{
    Probability density of the end-to-end distance for polymers in an active nematic background.  
    (a–c) Distributions for polymers of increasing length at selected activity levels ($\zeta = 0.0$, $0.03$, $0.05$, $0.08$).
    }
    \label{fig:pdf}
\end{figure}

To further explore the impact of activity on polymer conformations, we analyze the probability density of the end-to-end distance for polymers of different lengths. 
For polymer with $N=100$ monomers, the mean end-to-end distance consistently exceeds that of the passive state, and the broad PDFs indicate that the polymer explores a significantly wider conformational space (\fig{fig:pdf}a).
For a polymer with $N = 10$ monomers, the mean shifts toward smaller values, indicating compaction (\fig{fig:pdf}b).
Additionally, the PDFs maintain a roughly constant width, 
meaning that although the typical conformation shifts, the range of accessible conformations remains essentially unchanged.
For an intermediate chain length ($N = 25$), the PDFs initially shift toward lower values at low activity $\zeta<0.04$; however, with further increases in activity, the trend reverses—the mean shifts toward larger distances, indicating a transition to swelling.
The distributions also broaden, reflecting a more diverse range of conformations (\fig{fig:pdf}c). 
Flexible polymers undergo a collapse–stretch transition.

%%%%%%%%%%%%%%%%%%%%%%%%%%%%%%%%%%%%%%%%%%%%%%%%%%%%%%%%%%%%%
\section*{Zig-zag conformations and polymer compaction}
\label{appendix:More-Stiff}
%%%%%%%%%%%%%%%%%%%%%%%%%%%%%%%%%%%%%%%%%%%%%%%%%%%%%%%%%%%%%
\begin{figure}
    \centering
    \includegraphics[width=0.85\linewidth]{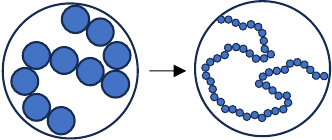}
    \caption{Schematic illustrating the refinement of polymer resolution.
    A coarse-grained polymer with $N=10$ large monomers (left) is replaced by a higher-resolution chain with $N=50$ smaller monomers (right), preserving the overall size (radius of gyration).
    }
    \label{fig:Moreschem}
\end{figure}

The observed zig-zag conformations—characterized by sharp bends between adjacent bonds—raise two key questions. 
First, are these zig-zags a genuine physical response to activity, or do they arise as artifacts of the chain's discretizations, given that they occur at the scale of individual monomers? In particular, since the zig-zag geometry is sensitive to steric interactions at the monomer level, we consider whether the excluded volume of individual beads artificially induces sharp local bending. 
Second, are these local distortions the primary drivers of the overall shrinkage observed in short polymers under active turbulence?

\begin{figure}
    \centering
    \includegraphics[width=\linewidth]{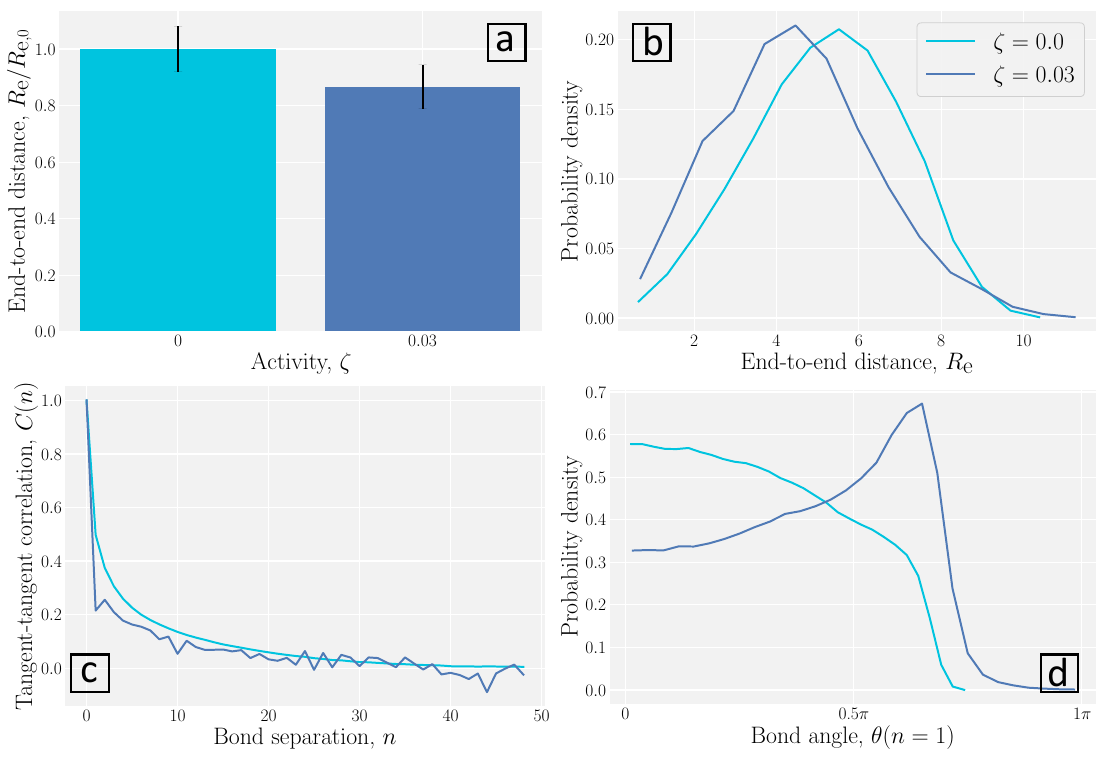}
    \caption{Effect of increased resolution on conformation.
    Conformational statistics for a polymer with $N=50$ monomers and shorter bond lengths $b=0.3$, chosen so that its passive-state size matches that of a standard $N = 10$ with $b=0.9$.
    (a) Time-ensemble-averaged end-to-end distance $\re/\reTrans$. (b) Probability density of end-to-end distance. (c) Tangent–tangent correlation function $C(n)$. (d) Bond angle distribution.
    }
    \label{fig:MoreMon}
\end{figure}
To address the first question, we decrease the thickness of the chain by reducing the monomer size—without changing its overall size, thereby relaxing local steric constraints while preserving global conformation (\fig{fig:Moreschem}).
Specifically, we simulate a polymer with $N=50$ monomers and shorter bond lengths of $b = 0.3$, set by adjusting the WCA diameter to $\sigma = 0.3$ (\eq{eq:WCA}) and the FENE maximum bond length to $r_0 = 0.45$ (\eq{eq:FENE}).
These parameters are chosen such that its passive-state radius of gyration matches that of the $N=10$ polymer studied in \fig{fig:correlations}.
The high-resolution, thinner chain also exhibits shrinkage at activity $\zeta=0.03$ (\fig{fig:MoreMon}a,b), similar to the compaction observed for the standard $N=10$ chain (\fig{fig:pdf}). 
The tangent–tangent correlation function shows a sharp drop at the nearest-neighbor separation $C(1)$, followed by a slight increase at the next-nearest neighbor $C(2)$ (\fig{fig:MoreMon}c).
While this behavior qualitatively resembles that of the standard $N=10$ chain with larger monomers (\fig{fig:correlations}a), a key difference emerges: the sharp bend now occurs over a shorter physical distance ($b=0.3$), reflecting the reduced monomer size.
This indicates that the zig-zag conformation is not governed by a fixed physical bending length, but instead occurs between adjacent monomers—regardless of their size.

\begin{figure}
    \centering
    \includegraphics[width=\linewidth]{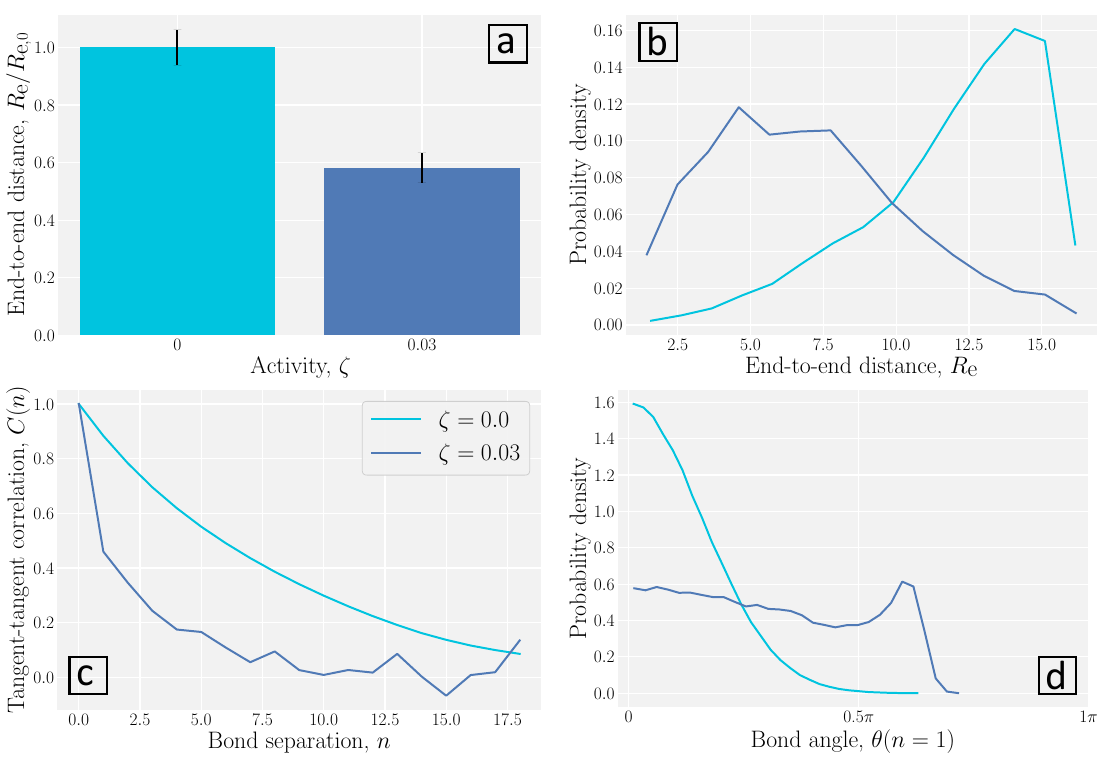}
    \caption{Effect of increased stiffness on conformation.
    Conformational statistics for a polymer with $N=20$ monomers and stiffer bonds, using the same bead size as in \fig{fig:correlations}.
    (a) Time-ensemble-averaged end-to-end distance $\re/\reTrans$. (b) Probability density of end-to-end distance. (c) Tangent–tangent correlation function $C(n)$. (d) Bond angle distribution.
    }
    \label{fig:Stiff}
\end{figure}
Having established that the zig-zag conformation is closely tied to the discretizations of the polymer model, we next turn to the second question: are these local distortions the primary drivers of shrinkage in short polymers? In other words, if we suppress sharp bending between adjacent monomers, does the polymer still undergo compaction under activity? To test this, we simulate a polymer with increased backbone stiffness, designed to resist bond-level deformation while preserving the same monomer size as in the standard model.
Specifically, we simulate a polymer with $N=20$ monomers and increased backbone stiffness, while keeping the same monomer size as in \fig{fig:REE}.
The stiffness is implemented via an angular potential \( V_{\text{bend}} = \frac{k_{\text{bend}}}{2}(\theta_{ijk} - \vartheta_{ijk})^2 \), which penalizes deviations from a straight configuration (with \( k_{\text{bend}} = 4 \) and \( \vartheta_{ijk} = 0 \)), thereby suppressing sharp local bending.
Despite this constraint, the stiff polymer still exhibits a notable reduction in its end-to-end distance under activity (\fig{fig:Stiff}a,b), indicating that compaction persists even in the absence of bond-level flexibility. 
Its tangent–tangent correlation function decays gradually, without the abrupt drop at $C(1)$ followed by a slight increase at the next-nearest neighbor $C(2)$ characteristic of zig-zag conformations (\fig{fig:Stiff}c).
The bond angle distribution reflects only moderate bending, with no significant weight at large angles (\fig{fig:Stiff}d).

These results demonstrate that the sharp local decorrelation observed in flexible chains is not required for activity-induced shrinkage.
Instead, they suggest that compaction arises from global conformational folding driven by the active environment, independent of local zig-zag geometry.
We thus conclude that, while the specific zig-zag conformation is related to the finite-size of the MD monomers, the compaction of short polymers is physical.

%%%%%%%%%%%%%%%%%%%%%%%%%%%%%%%%%%%%%%%%%%%%%%%%%%%%%%%%%%%%%
%%%%%%%%%%%%%%%%%%%%%%%%%%%%%%%%%%%%%%%%%%%%%%%%%%%%%%%%%%%%%
%% END OF MAIN TEXT
%%%%%%%%%%%%%%%%%%%%%%%%%%%%%%%%%%%%%%%%%%%%%%%%%%%%%%%%%%%%%
%%%%%%%%%%%%%%%%%%%%%%%%%%%%%%%%%%%%%%%%%%%%%%%%%%%%%%%%%%%%%

%%%%%%%%%%%%%%%%%%%%%%%%%%%%%%%%%%%%%%%%%%%%%
%%%%%%%%%%%%%%%%%%%%%%%%%%%%%%%%%%%%%%%%%%%%%
%%%%%%%%%%%%%%%%%%%%%%%%%%%%%%%%%%%%%%%%%%%%%
%% Bibliography
%%%%%%%%%%%%%%%%%%%%%%%%%%%%%%%%%%%%%%%%%%%%%
%%%%%%%%%%%%%%%%%%%%%%%%%%%%%%%%%%%%%%%%%%%%%
%%%%%%%%%%%%%%%%%%%%%%%%%%%%%%%%%%%%%%%%%%%%%
% \bibliographystyle{unsrt}
% \bibliography{references}
\input{ConformationMain.bbl}

\end{document}